\documentstyle[12pt]{article}

\input epsf.tex

\def\ll{\lambda\lambda}
\def\vev#1{\mathord < #1 \mathord >}

\def\Sb{\bar{S}}
\def\Tb{\bar{T}}

\def\beq{\begin{equation}}
\def\eeq{\end{equation}}
\def\TR{T+\Tb-Y\Yb}

\newcommand{\req}[1]{(\ref{#1})}

\newcommand{\nwc}{\newcommand}
\nwc{\btu}{\bigtriangleup}
\nwc{\cd}{\cdot}
\nwc{\zd}{{\bf Z}$_3$\ }
\nwc{\hyp} {\hyphenation}
\hyp{orbi-fold} 
\hyp{theo-ries}
\hyp{theo-ry}
\hyp{regu-lari-zation}
\newcommand{\Z}{\ZZ}
\def\bfone{\relax{\rm 1\kern-.35em 1}}
\def\inbar{\vrule height1.5ex width.4pt depth0pt}
\def\IC{\relax\,\hbox{$\inbar\kern-.3em{\mss C}$}}
\def\ID{\relax{\rm I\kern-.18em D}}
\def\IF{\relax{\rm I\kern-.18em F}}
\def\IH{\relax{\rm I\kern-.18em H}}
\def\II{\relax{\rm I\kern-.17em I}}
\def\IN{\relax{\rm I\kern-.18em N}}
\def\IP{\relax{\rm I\kern-.18em P}}
\def\IQ{\relax\,\hbox{$\inbar\kern-.3em{\rm Q}$}}
\def\IR{\relax{\rm I\kern-.18em R}}
\def\ZZ{\relax{\hbox{\mss Z\kern-.42em Z}}}
\font\cmss=cmss10 \font\cmsss=cmss10 at 7pt
\def\ZZ{\relax\ifmmode\mathchoice
{\hbox{\cmss Z\kern-.4em Z}}{\hbox{\cmss Z\kern-.4em Z}}
{\lower.9pt\hbox{\cmsss Z\kern-.4em Z}}
{\lower1.2pt\hbox{\cmsss Z\kern-.4em Z}}\else{\cmss Z\kern-.4em Z}\fi}
\nwc{\ten}{ten--dimensional}
\nwc{\four}{four--dimensional}
\nwc{\gev} {{\rm GeV}}
\nwc{\tev} {{\rm TeV}}
\nwc{\mP} {$M_{\rm Planck}$}
\nwc{\mx} {$M_{\rm X}$}
\nwc{\ms} {$M_{\rm string}$}
\nwc{\sieb}{\mbox{\boldmath $\ov{27}$}}
\nwc{\sie}{\mbox{\boldmath ${27}$}}
\nwc{\lag}{Lagrangian}

\newcommand{\tcgc}{threshold corrections to the gauge couplings}
\newcommand{\tc}{threshold corrections}

\nwc{\be}  {\begin{equation}}
\nwc{\ee}  {\end{equation}}
\nwc{\ba}  {\begin{array}}
\nwc{\ea}  {\end{array}}
\nwc{\bdm} {\begin{displaymath}}
\nwc{\edm} {\end{displaymath}}
\nwc{\bea} {\be\ba{lcl}}
\nwc{\eea} {\ea\ee}
\nwc{\bda} {\bdm\ba{lcl}} 
\nwc{\eda} {\ea\edm}
\nwc{\bc}  {\begin{center}}
\nwc{\ec}  {\end{center}}
\nwc{\ds}  {\displaystyle}
\nwc{\bmat}{\left(\ba}
\nwc{\emat}{\ea\right)}
\nwc{\nn} {\nonumber}
\nwc{\nnn} {\nonumber \vspace{.2cm} \\ }
\nwc{\ra}{\rightarrow}
\nwc{\lra}{\longrightarrow}
\nwc{\p} {\partial}

\nwc{\scr}  {\scriptstyle}
\nwc{\tx}  {\textstyle}
\nwc{\scs} {\scriptscriptstyle}
\nwc{\ov}  {\overline}
\nwc{\hb}  {\bar h}
\nwc{\xb}  {\bar x}
\nwc{\yb}  {\bar y}
\nwc{\zb}  {\bar z}
\nwc{\wb}  {\bar w}
\nwc{\Ob}  {\bar O}
\nwc{\Yb}  {\bar Y}
\nwc{\ep} {\epsilon}
\nwc{\de} {\delta}
\nwc{\Th} {\Theta}
\nwc{\th} {\theta}
\nwc{\al} {\alpha}
\nwc{\si} {\sigma}
\nwc{\Si} {\Sigma}
\nwc{\om} {\omega}
\nwc{\Om} {\Omega}
\nwc{\Ga} {\Gamma}
\nwc{\ga} {\gamma}
\nwc{\bet} {\beta}
\nwc{\La} {\Lambda}
\nwc{\la} {\lambda}
\nwc{\Sc}  {{\cal S}}
\nwc{\Rc}  {{\cal R}}
\nwc{\Dc}  {{\cal D}}
\nwc{\Oc}  {{\cal O}}
\nwc{\Cc}  {{\cal C}}
\nwc{\gc}  {{\cal g}}
\nwc{\Pc}  {{\cal P}}
\nwc{\Mc}  {{\cal M}}
\nwc{\Ec}  {{\cal E}}
\nwc{\Fc}  {{\cal F}}
\nwc{\Hc}  {{\cal H}}
\nwc{\Kc}  {{\cal K}}
\nwc{\Wc}  {{\cal W}}

\nwc{\Fcp} {{\cal F}^\pr}
\nwc{\Hcp} {{\cal H}^\pr}

\nwc{\Xc}  {{\cal X}}
\nwc{\Gc}  {{\cal G}}
\nwc{\Zc}  {{\cal Z}}
\nwc{\Nc}  {{\cal N}}

\nwc{\xc}  {{\cal x}}
\nwc{\Ac}  {{\cal A}}
\nwc{\Bc}  {{\cal B}}
\nwc{\Uc} {{\cal U}}
\nwc{\Vc} {{\cal V}}
\nwc{\Lc} {{\cal L}}
\nwc{\Qc} {{\cal Q}}

\nwc{\lng} {\langle}
\nwc{\rng} {\rangle}

\nwc{\lf} {\left}
\nwc{\ri} {\right}

\nwc{\diag} {{\rm diag}}
\nwc{\inv}  {{\rm inv}}
\nwc{\mod}  {{\ \rm mod\ }}
\nwc{\dete}  {{\rm det}}
\nwc{\tr}  {{\rm tr}}
\nwc{\im}  {{\rm Im}}
\nwc{\re}  {{\rm Re}}

\nwc{\h} {\frac{1}{2}}
\nwc{\fc} {\frac}

\def\KK{\relax{\rm I\kern-.18em K}}
\def\RR{\relax{\rm I\kern-.18em R}}
\def\NN{\relax{\rm I\kern-.18em N}}
\def\PP{\relax{\rm I\kern-.18em P}}

\def\zz{\relax{\sf Z\kern-.3em Z}}
\def\ZZ{\relax{\sf Z\kern-.4em Z}}
\def\ZZZ{{\relax{\sf Z}\kern -.5em Z}}
\def\ZZZ{Z\kern -0.37em Z}
\def\QQ{{\rm \kern .25em
             \vrule height1.4ex depth-.12ex width.06em\kern-.31em Q}}
\def\CC{{\rm \kern .25em
             \vrule height1.4ex depth-.12ex width.06em\kern-.31em C}}
\begin{document}

\begin{titlepage}
{\sf
\begin{flushright}
{TUM--HEP--234/96}\\
{SFB--375/28}\\
{January 1996}
\end{flushright}}
\vfill
\begin{center}
{\large \bf Dynamical Gauge Coupling Constants$^{\mbox{\boldmath $\ast$}}$}

\vskip 1.2cm
{\sc Hans Peter Nilles}\\
%
\vskip 1.5cm
{\em Institut f\"{u}r Theoretische Physik} \\
{\em Physik Department T30} \\
{\em Technische Universit\"at M\"unchen} \\
{\em D--85747 Garching} \\
{\em Germany}
\vskip 1cm
{\em Max--Planck--Institut f\"ur Physik} \\
{\em ---Werner--Heisenberg--Institut---}\\
{\em P.O. Box 401212}\\
{\em D--80805 M\"{u}nchen} \\
{\em Germany}
\end{center}
\vfill

\thispagestyle{empty}

\begin{abstract}
In string theory  the coupling parameters are functions of moduli fields.
The actual values of the coupling constants are then dynamically 
determined through the vacuum expextation values of these fields. We
review the attempts to connect such theories to low energy effective
field theories with realistic gauge coupling constants. This includes
a discussion of supersymmetry breakdown, the question of a running
dilaton, string threshold calculations and the possibility of string
unification.
\end{abstract}

\vskip 5mm \vskip0.5cm
\hrule width 5.cm \vskip 1.mm
{\small\small  $^\ast$ Lectures given at the 1995 Trieste Summer School
in High Energy Physics and Cosmology, Trieste, Italy, June 1995.}
\end{titlepage}

\section{Introduction}
Most physics models contain coupling constants as free parameters that
can be adjusted to fit the experimental values. In  more complete
theories one could imagine that the values of these parameters are
determined dynamically by the theory itself. The question arises, how and
why the coupling parameters take the values that are observed in 
nature.

String theory\cite{gsw} provides an example for such a class of models. 
Couplings are functions  of
so-called moduli fields and the actual values of the coupling parameters 
are determined through the vacuum expectation values (vev) of these fields.
At tree level the universal gauge coupling constant $g_{\rm string}$ is 
determined by
the vev of the dilaton field\cite{10} 

\beq
S= {4\pi\over{g^2_{\rm string}}} + i {\theta\over{2\pi}}.
\eeq
Nonuniversalities can appear at the one-loop level and depend on further
moduli fields $T$, $U$ or $B$ 

\be
\fc{1}{g_a^2(\mu)}=\fc{k_a}{g_{\rm string}^2}+\fc{b_a}{16\pi^2} 
\ln \fc{M_{\rm string}^2}{\mu^2}-\fc{1}{16\pi^2}\triangle(T, U, B\ \ldots )\ .
\ee
Given this situation we have then to see how a theory with a realistic
set of gauge coupling constants can emerge.

We would therefore like to connect such a theory with a low energy
effective field theory describing the known particle physics phenomena.
A prime candidate is a low energy supergravitational
generalization of the standard model of strong and electroweak
interactions, with supersymmetry broken in a hidden sector\cite{sugra}.
Unification of observable sector gauge couplings might appear as proposed
in supersymmetric grand unified theories (SUSYGUTs)\cite{guts}.
The vevs of the moduli fields $S$, $T$, $\ldots$ should then determine
gauge couplings in hidden and observable sector, including the correct
values of the QCD coupling $\alpha_s$ and the weak mixing angle
$\sin^2 \theta_{\rm W}$.

In string theories with unbroken supersymmetry the vevs of the
moduli fields are undetermined. A first step in the determination 
of gauge coupling constants requires therefore the discussion of 
supersymmetry breakdown. We then have to see how the vevs of the
moduli are fixed. Of course, not any value of the moduli will lead to
a satisfactory model. In fact we shall face some generic problems
concerning the actual values of the coupling constants. We have to
understand why the value of

\beq
\alpha_{\rm string}={g^2_{\rm string}\over{4\pi}}
\approx {1\over{20}} ,
\eeq
whereas a natural expectation for the vev of $S$ would be a number of
order 1 or maybe 0 or even infinity, as happens in many simple models.
A related question concerns the possibility of so-called string
unification leading to the correct prediction of the weak
mixing angle $\sin^2 \theta_{\rm W}$. Naively one would expect string
unification to appear at $M_{\rm string}\approx 4\times 10^{17}$GeV,
while the correct prediction of $\sin^2 \theta_{\rm W}$ seems to 
lead to a scale that is a factor of 20 smaller. We then have to face the
question how such a situation can be achieved with natural values of the
vevs of the moduli fields. 

These are the questions we want to address in these lectures. We shall
start with the discussion of supersymmetry breakdown in the framework
of gaugino condensation. This includes a discussion of the problem
of a "runaway" dilaton that any attempt of a dynamical determination of
coupling constants has to face. The next question then concerns the
value of $<S>\approx 1$ and its compatibility with weak coupling. 
Finally we shall discuss new results concerning string threshold
corrections and the question of string unification\cite{nist}.

\section{Gaugino Condensation}

One of the prime motivations to consider the supersymmetric extension of the
standard model is the stability of the weak scale ($M_W$) of order of a TeV in
the presence of larger mass scales like a GUT-scale of
$M_X\approx 10^{16}\;GeV$ or
the Planck scale $M_{Pl} \approx 10^{18}\;GeV$. The size of the weak scale is
directly related to the breakdown scale of supersymmetry, and a satisfactory
mechanism of supersymmetry breakdown should explain the smallness of
$M_W/M_{Pl}$ in a natural way. One such mechanism is based on the dynamical
formation of gaugino condensates that has attracted much attention since its
original proposal for a spontaneous breakdown of supergravity \cite{1}\cite{2}.
In the following  we shall address some open questions 
concerning this mechanism in
the framework of low energy effective superstring theories. This work has been
done in collaboration with Z. Lalak and 
A. Niemeyer and appeared in ref. \cite{3}\cite{4}.

Before discussing these detailed questions let us remind you of the basic facts
of this mechanism. For simplicity we shall consider here a pure supersymmetric
($N=1$) Yang-Mills theory, with the vector multiplet $(A_\mu, \lambda)$
containing gauge bosons and gauge fermions in the adjoint representation of
the nonabelian gauge group. Such a theory is asymptotically free and we would
therefore (in analogy to QCD) expect confinement and gaugino condensation at
low energies \cite{5}. We are then faced with the question whether such a
nontrivial gaugino condensate $\mathord < \lambda\lambda \mathord > \neq 0$
leads to a breakdown of supersymmetry. A first look at the
SUSY-transformation on the composite fermion $\lambda\sigma^\mu A_\mu$
\cite{dfs} 

\beq
\{Q,  \lambda\sigma^\mu A_\mu\}=\lambda\lambda + \ldots
\eeq

\noindent might suggest a positive answer, but a careful inspection of the
multiplet 
structure and gauge invariance leads to the opposite conclusion. The bilinear
$\lambda\lambda$ has to be interpreted as the lowest component of the chiral
superfield 
$W^\alpha W_\alpha=(\ll, \ldots)$ and therefore a non-vanishing vev of $\ll$
does not break SUSY \cite{6}. This suggestion is supported by index-arguments
\cite{7} and an effective Lagrangian approach \cite{8}. We are thus lead to the
conclusion that in such theories gaugino condensates form, but do not break
global (rigid) supersymmetry.

Not all is lost, however, since we are primarily interested in models with
local supersymmetry including gravitational interactions. The weak
gravitational force should not interfere with the formation of the condensate;
we therefore still assume  $\mathord < \lambda\lambda \mathord > = \Lambda^3
\neq 0$. This expectation is confirmed by the explicit consideration of the
effective Lagrangian of ref. \cite{1} in the now locally supersymmetric
framework. We here consider a composite chiral superfield $U=(u, \psi, F_u)$
with $u= \mathord < \lambda\lambda \mathord >$. In this toy model
\cite{1}\cite{2} we obtain the surprising result that not only $\vev u =
\Lambda^3 \neq 0$ but also $\vev{F_u} \neq 0$, a signal for supersymmetry
breakdown. In fact

\beq
\vev{F_u} = M_S^2 = \frac{\Lambda^3}{M_{Pl}},
\eeq

\noindent consistent with our previous result that in the global limit $M_{Pl}
\rightarrow \infty$ (rigid) supersymmetry is restored. For a hidden sector
supergravity model we would choose $M_S \approx 10^{11}\;GeV$ \cite{2}.

Still more information can be obtained by consulting the general supergravity
Lagrangian of elementary fields determined by the K\"ahler potential $K(\Phi_i,
{\Phi^j}^\ast)$, the superpotential $W(\Phi_i)$ and the gauge kinetic function
$f(\Phi_i)$ for a set of chiral superfields $\Phi_i=(\phi_i, \psi_i, F_i)$.
Non-vanishing vevs of the auxiliary fields $F_i$ would signal a breakdown of
supersymmetry. In standard supergravity notation these fields are given by

\beq
F_i = \exp(G/2) (G^{-1})^j_i G_j + \frac 14 \frac{\partial f}{\partial \Phi_k}
(G^{-1})^k_i \ll + \ldots ,\label{eq3}
\eeq

\noindent where the gaugino bilinear appears in the second term \cite{9}. This
confirms 
our previous argument that $\vev \ll \neq 0$ leads to a breakdown of
supersymmetry, however, we obtain a new condition: $\partial f/\partial \Phi_i$
has to be nonzero, i.e. the gauge kinetic function $f(\Phi_i)$ has to be
nontrivial. In the fundamental action $f(\Phi_i)$ multiplies $W_\alpha
W^\alpha$ which in components leads to a form $\mbox{Re} f(\phi_i) F_{\mu\nu}
F^{\mu\nu}$ and tells us that the gauge coupling is field dependent. For
simplicity we consider here one modulus field $M$ with 

\beq
\vev {\mbox{Re} f(M)} \approx 1 /g^2.
\eeq

This dependence of $f$ on the modulus $M$ is very crucial for SUSY breakdown
via gaugino condensation. $\partial f/\partial M \neq 0$ leads to $F_M\approx
\Lambda^3/M_{Pl}$ consistent with previous considerations. The goldstino is the
fermion in the $f(M)$ supermultiplet. In the full description of the theory it
might mix with a composite field, but the inclusion of the composite fields
should not alter the qualitative behaviour discussed here. An understanding of
the mechanism of SUSY breakdown via gaugino condensation is intimately related
to the question of a dynamical determination of the gauge coupling constant as
the vev of a modulus field. We would hope that in a more complete theory such
questions could be clarified in detail.

One candidate of such a theory is the $E_8 \times E_8$ heterotic string. The
second $E_8$ (or a subgroup thereof) could serve as the hidden sector gauge
group and it was soon found \cite{10} that there we have nontrivial $f=S$ where
$S$ represents the dilaton superfield. The heterotic string thus contains all
the necessary ingredients for a successful implementation of the mechanism of
SUSY breakdown via gaugino condensation \cite{11}\cite{12}. Also the question
of the dynamical determination of the gauge coupling constant can be addressed.
A simple reduction and truncation from the $d=10$ theory leads to the following
scalar potential \cite{13}

\beq
V=\frac 1{16 S_R T_R^3} \left [ |W(\Phi) + 2 (S_R T_R)^{3/2} (\ll)|^2 + \frac
{T_R}3 \left |\frac{\partial W}{\partial \Phi} \right |^2 \right]\label{eq5},
\eeq

\noindent where $S_R=\mbox{Re} S$, $T_R=\mbox{Re} T$ is the modulus
corresponding 
to the overall radius of compactification and $W(\Phi)$ is the superpotential
depending on the matter fields $\Phi$. The gaugino bilinear appears via the
second term in the auxiliary fields (\ref{eq3}). To make contact with the
dilaton field, observe that $\vev \ll = \Lambda^3$ where $\Lambda$ is the
renormalization group invariant scale of the nonabelian gauge theory under
consideration. In the one-loop approximation

\beq
\Lambda = \mu \exp \left ( -\frac 1{bg^2(\mu)}\right), 
\eeq

\noindent with an arbitrary scale $\mu$ and the $\beta$-function coefficient
$b$. This then suggests

\beq
\ll \approx e^{-f} = e^{-S} \label{eq7}
\eeq

\noindent as the leading contribution (for weak coupling) for the functional 
$f$-dependence of the gaugino bilinear\footnote{ Relation (\ref{eq7}) is of
course not exact. For different implementations see \cite{12}, \cite{14},
\cite{15}. The qualitative behaviour of the potential remains unchanged.}.

In the potential (\ref{eq5}) we can then insert (\ref{eq7}) and determine the
minimum. In our simple model (with $\partial W/\partial T=0$) we have a
positive definite potential with vacuum energy $E_{vac}=0$. Suppose now for the
moment that $\vev{W(\Phi)} \neq 0$\footnote{In many places in the literature it
is quoted incorrectly that  $\vev{W(\Phi)}$ is quantized in units of the Planck
length since $W$ comes from $H$, the field strength of the antisymmetric tensor
field $B$ and $H=\mbox{d}B -\omega_{3Y} + \omega_{3L}$ ($\omega$ being the
Chern-Simons form). Quantization is
expected for $\vev{\mbox{d}B}$ but not necessarily for $H$.}.

\noindent $S$ will now adjust its vev in such a way that $|W(\Phi) + 2 (S_R
T_R)^{3/2} (\ll)| =0$, thus

\beq
|W(\Phi) + 2 (S_R T_R)^{3/2} \exp(-S)| = 0.
\eeq

\noindent This then leads to
broken SUSY with $E_{vac} =0$ and a fixed value of the gauge coupling constant
$g^2 \approx \vev{\mbox{Re} S}^{-1}$. For the vevs of the auxiliary fields we
obtain $F_S=0$ and $F_T\neq 0$ with important consequences for the pattern of
the soft SUSY breaking terms in phenomenologically oriented models \cite{brig},
which we shall not discuss here in detail.

Thus a satisfactory picture seems to emerge. However, we have just discussed a
simplified example. In general we would expect also that the superpotential
depends on the moduli, $\partial W/\partial T\neq 0$ and, including this
dependence, the modified potential would no longer be positive definite and one
would have  $E_{vac}<0$. 

But even in the simple case we have a further vacuum degeneracy. For any value
of $W(\Phi)$ we obtain a minimum with $E_{vac}=0$, including $W(\Phi)=0$. In
the latter case this would correspond to $\vev \ll=0$ and $S\rightarrow
\infty$. This is the potential problem of the runaway dilaton. The simple model
above does not exclude such a possibility. In fact this problem of the runaway
dilaton does not seem just to be a problem of the toy model, but more
general. One attempt to avoid this problem was the consideration of several
gaugino 
condensates \cite{kraslal}, but it still seems very difficult to produce
satisfactory potentials that lead to a dynamical determination of the dilaton
for reasonable values of $\vev S$. In some cases it even seems impossible to
fine tune the cosmological constant to zero. In absence of a completely
satisfactory model it is then also difficult to investigate the detailed
phenomenological properties of the approach. Here it would be of interest to
know the actual size of the vevs of the auxiliary fields $\vev{F_S}$,
$\vev{F_T}$ and $\vev{F_U}$. In the models discussed so far one usually finds
$\vev{F_T}$ to be the dominant term, but it still remains a question whether
this is true in general.

\section{Fixing the dilaton}

In any case it seems that we need some new ingredient before we can understand
the mechanism completely. It is our belief, that the resolution of all these
problems comes with a better understanding of the form of the gauge kinetic
function $f$ \cite{3}\cite{4}. In all the previous considerations one assumed
$f=S$. How general is this relation? Certainly we know that in one loop
perturbation theory $S$ mixes with $T$ \cite{stmix}, but this is not
relevant for our discussion and, for simplicity, we shall ignore that for the
moment. The formal relation between $f$ and the condensate is given through
$\Lambda^3 \approx e^{-f}$ and we have $f=S$ in the weak coupling limit of
string theory. In fact this argument tells us only that

\beq
\lim_{S\rightarrow \infty} f(S) = S\label{eq8}.
\eeq 
Nonperturbative effects could lead to the situation that $f$ is a very
complicated function of $S$. In fact a satisfactory incorporation of gaugino
condensates in the framework of string theory might very well lead to such a
complication. In our work \cite{3} we suggested that a nontrivial $f$-function
is the key ingredient to better understand the mechanism of gaugino
condensation. We still assume (\ref{eq8}) to make contact with perturbation
theory. How do we then control $e^{-f}$ as a function of $S$? In absence of a
determination of $f(S)$ by a direct calculation one might use symmetry
arguments to make some progress. Let us here consider the presence of a
symmetry called $S$-duality which in its simplest form is given by a $SL(2,Z)$
generated by the transformations

\beq
S\rightarrow S+i,\quad S\rightarrow -1/S.
\eeq

Such a symmetry might be realized in two basically distinct ways: 
the gauge sector could close under the transformation (type I) or being mapped
to an additional `magnetic sector' with inverted coupling constant (type II).
In the second
case one would speak of strong-weak coupling duality, just as in the case of
electric-magnetic duality \cite{seiwi}. Within the class of theories of type I,
however, we 
could have the situation that the $f$-function is itself 
invariant\footnote{More complicated choices of transformation properties
for $f$ are possible and lead to similar results as obtained in our
simple toy model.} under
$S$-duality; i.e. $S\rightarrow -1/S$ does not invert the coupling constant
since the gauge coupling constant is not given by $\mbox{Re} S$ but $1/g^2
\approx \mbox{Re} f$. In view of (\ref{eq8}) we would call such a symmetry
weak-weak coupling duality. The behaviour of the gauge coupling constant as a
function of $S$ is shown in Fig. 1. Our assumption (\ref{eq8}) implies that
$g^2 \rightarrow 0$ as $\mbox{Re} S \rightarrow \infty$ and by $S$-duality
$g^2$ also vanishes for $S\rightarrow 0$, with a maximum somewhere in the
vicinity of the self-dual point $S=1$. Observe that $S\approx 1$ in this
situation does not necessarily imply strong coupling, because  $g^2
\approx 1/\mbox{Re} f$ and even for $S\approx 1$, $\mbox{Re} f$ could be large
and $g^2<<1$, with perturbation theory valid in the whole range of $S$. Of
course, nonperturbative effects are responsible for the actual form of $f(S)$.

\epsfbox[-80 0 500 210]{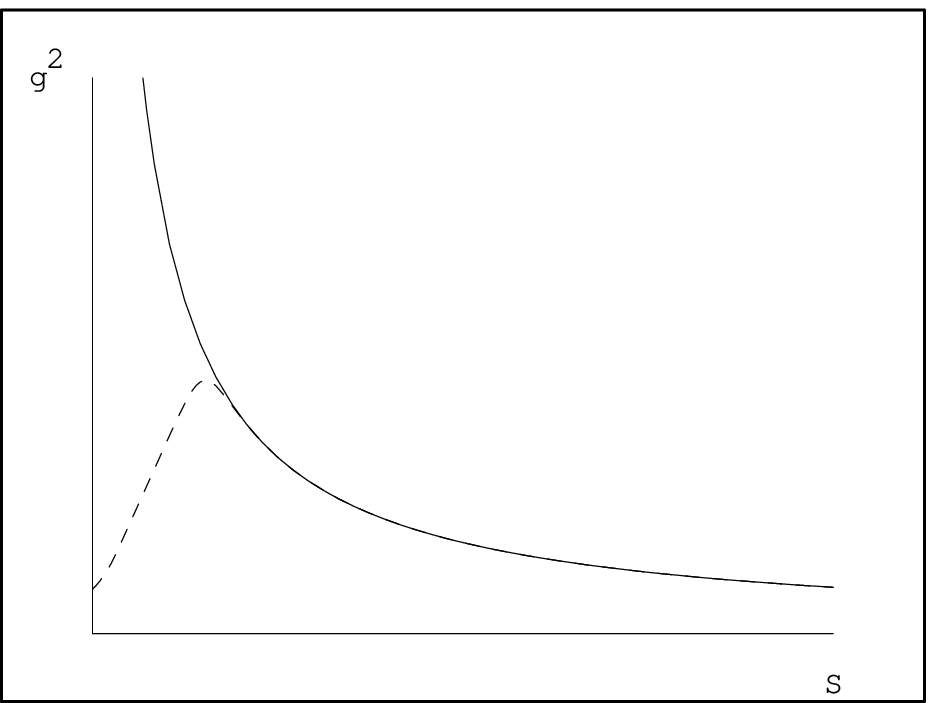}

{\small \em \noindent Fig. 1 - Coupling constant $g^2$ as
the function of $S$ in type-I models (dashed) vs $g^2$ given by $f=S$}

\vspace{0.3cm}

To examine the behaviour of the scalar potential in this approach, let us
consider a simple toy model, with chiral superfields 
$U=Y^3 = (\ll, \ldots )$ as
well as $S$ and $T$. We have to choose a specific example of a gauge kinetic
function which is invariant under the $S$-duality transformations. Different
choices are possible, the simplest is given by

\beq
f=\frac 1{2\pi} \ln(j(S)-744),
\eeq

\noindent $j(S)$ being the usual generator of modular invariant functions. This
function behaves like $S$ in the large $S$-limit. If we assume a type I-model
where the 
gauge sector is closed under $S$-duality, then
we also have to assume that the gaugino condensate does not transform under
$S$-duality (because of the $f W^\alpha W_\alpha$-term in the
Lagrangian)\footnote{For type I-models it was shown in \cite{3} that one can
always redefine the gauge kinetic function and condensate in such a way that
this holds.}. Under these conditions an obvious candidate for the
superpotential is just the standard Veneziano-Yankielowicz superpotential
(extended to take into account the usual $T$-duality, which we assume to be
completely independent from $S$-duality) \cite{8}\cite{tdual}

\beq
W=Y^3 (f + 3b\ln \frac {Y\eta^2(T)}{\mu} +c).
\eeq

This is clearly invariant under $S$-duality. Therefore we then cannot take the
conventional form for the K\"ahler potential which would be given by

\beq
K= -\ln(S+\Sb)-3\ln(\TR),
\eeq

\noindent since it is not $S$-dual. To make it $S$-dual one could introduce an
additional $\ln |\eta(S)|^4$ term, giving e.g.

\beq
K=-\ln(S+\Sb)-3\ln(\TR)-\ln |\eta(S)|^4.
\eeq

\noindent Because the only relevant quantity is

\beq
G=K+\ln |W|^2,
\eeq

\noindent we can as well put this new term (which is forced upon us because of
our demand 
for symmetry) into the superpotential and take the canonical K\"ahler function
instead, which gives

\beq
K=-\ln(S+\Sb)-3\ln(\TR),
\eeq
\beq
W=\frac{Y^3}{\eta^2(S)} (f + 3b\ln \frac {Y\eta^2(T)}{\mu} +c),
\eeq

\noindent where the remarkable similarity to the effective potential for
$T$-dual gaugino condensation $W=W_{inv}/\eta^6(T)$ can be seen more clearly.

This model exhibits a well defined minimum at $\vev S =1$, $\vev T=1.23$ and
$\vev Y\approx \mu$. Supersymmetry is broken with the dominant contribution
being $\vev{F_T}\approx \mu^3$. The cosmological constant is negative.

In contrast to earlier attempts \cite{ccm} this model fixes the
problem of the runaway dilaton and breaks supersymmetry with only a single
gaugino condensate. Previous models needed multiple gaugino condensates and (to
get realistic vevs for the dilaton) matter fields in complicated
representations. We feel that the concept of a nontrivial gauge kinetic
function derived (or constrained) by a symmetry is a much more natural way to
fix the dilaton and break supersymmetry, especially so because corrections to
$f=S$ are expected in any case. Earlier models which included $S$-duality in
different ways (both with and without gaugino condensates)
\cite{filq}\cite{hm}
were able to fix the vev of the dilaton but did not succeed in breaking
supersymmetry. An alternative mechanism to fix the vev of the dilaton has been
discussed in \cite{macross}.

Of course there are still some open questions not solved by this approach. The
first is the problem of having a vanishing cosmological constant. Whereas early
models of gaugino condensation often introduced {\em ad hoc} terms to guarantee
a vanishing vacuum energy, it has been seen to be notoriously difficult to get
this out of models based on string inspired supergravity. The only way out of
this problem so far has been to introduce a constant term into the
superpotential, parameterizing unknown effects. This approach does not even
work in any arbitrary model, but at least in our model the cosmological
constant can be made to vanish by adjusting such a constant.

Another question not addressed in this toy model is the mixing of $S$ and $T$
fields which happens at the one-loop level. It is still unknown whether one can
keep two independent dualities in this case. In a consistent interpretation our
toy model should describe an all-loop effective action. If it is considered to
be a theory at the tree-level then the 
theory is not anomaly free. Introducing terms to cancel the anomaly
which arises because of demanding $S$-duality will then destroy $S$-duality. At
tree-level the theory therefore cannot be made anomaly free.

An additional interesting question concerns the vevs of the
auxiliary fields, i.e.  which field is responsible for
supersymmetry breakdown. 
In all models considered so far (multiple gaugino condensates, matter,
S-duality) it has always been $F_T$ which dominates all the other auxiliary
fields. It has not been shown yet that this is indeed a generic feature. The
question is an important one, since the hierarchy of the vevs of the auxiliary
fields is mirrored in the structure of the soft SUSY breaking terms of the
MSSM \cite{brig}. We want to argue that there is at least no evidence for $F_T$
being generically large in comparison to $F_S$, because all of the models
constructed so far (including our toy model) are designed in such a way that
$\vev{F_S}=0$ by construction at the minimum (at least at tree-level for the
other models). In fact, if one extends our model with a constant in the
superpotential (see above), then $\vev{F_S}$ increases with the constant (but
does not become as large as $\vev{F_T}$).

Of course there are still some assumptions we made by considering this toy
model. We assumed that there is weak coupling in the large $S$ limit which is
an assumption because the nonperturbative effects are unknown (at tree-level it
can be calculated that $f=S$). In addition it is clear that the standard form
we take for the K\"ahler potential does not include nonperturbative effects and
thus could be valid only in the weak coupling approximation (this is of course
related to our choice of the superpotential). Of course an equally valid
assumption would be that nonperturbative effects destroy the calculable
tree-level behaviour even in the weak coupling region. The model of
ref. \cite{filq} could be re-interpreted in that sense (they do not consider
gaugino condensates and the gauge kinetic function, but their $S$-dual scalar
potential goes to infinity for $S\rightarrow \infty$). We choose not to make
this assumption, because it is equivalent to the statement that the whole
perturbative framework developed so far in string theory is wrong. 
Again it should be emphasized here that the
$S$-duality considered is not a strong-weak coupling duality but a
weak-weak coupling duality. In type II-models one has a duality between strong
and weak coupling \cite{3}. A detailed discussion of such a scenario
would be desirable.

\section{S=1 and weak coupling}

A problem could be the actual size of the gauge coupling constant. If
$f=S$ and $\vev S=1$ then the large value of the gauge coupling constant does
not fit the 
low scale of gaugino condensation necessary for phenomenologically realistic
supersymmetry breaking ($10^{13}\,GeV$). However if $f=S$ only in the weak
coupling limit then one can have $\vev f >> 1$ and thus $g^2<< 1$ even in
the region $S=O(1)$. Therefore in our model $\vev S=1$ is consistent with the
demand for a small gauge coupling constant, whereas in models with $f=S$ a much
larger (and therefore more unnatural) $\vev S$ is needed.

To summarize we find that the choice of a nontrivial $f$-function (motivated by
a symmetry requirement) gives rise to a theory where supersymmetry breaking is
achieved by employing only a single gaugino condensate. The cosmological
constant turns out to be negative, but can be adjusted by a simple additional
constant in the superpotential. The vevs of all fields are at natural orders of
magnitude and due to the nontrivial gauge kinetic function the gauge coupling
constant can be made small enough to give a realistic picture.

Turning our attention to the observable sector we see that a small 
(grand unified) coupling constant is a necessity and the above mechanism
is required for a satisfactory description of the size af the observed
coupling constants like e.g. $\alpha_{\rm QCD}$. But this alone
might not be sufficient for a realistic model. String theory should
predict all low energy coupling constants correctly and should also
give the correct ratio of electroweak and strong coupling constants.

LEP and SLC high precision electroweak data give
for the minimal supersymmetric Standard Model (MSSM) 
with the lightest Higgs mass in the range $60 \gev < M_{\rm H}<150 \gev$

\be
\ba{rcl}
\sin^2 \hat \th_{\rm W}(M_{\rm Z})&=&0.2316\pm0.0003\\
\al_{em}(M_{\rm Z})^{-1}&=&127.9\pm0.1\\
\al_{\rm S}(M_{\rm Z})&=&0.12\pm0.01\\
m_t&=& 160^{+11+6}_{-12-5}\gev\ ,
\ea\label{exp}
\ee
for the central value $M_{\rm H}=M_{\rm Z}$ in the $\ov{ MS}$ scheme
\cite{langacker}.
This is in perfect agreement with the recent CDF/D0 measurements of $m_t$.
Taking the first three values as input parameters leads to gauge coupling
unification at $M_{\rm GUT}\sim 2\cdot 10^{16}\gev$ with $\al_{\rm GUT}\sim 
\fc{1}{26}$ and 
$M_{\rm SUSY}\sim 1 {\rm TeV}$ \cite{uni00,langacker}.
Slight modifications arise from light SUSY thresholds, i.e. the splitting 
of the sparticle mass spectrum,
the variation of the mass of the second Higgs doublet
and two--loop effects. Whereas these
effects are rather mild, huge corrections may arise from heavy thresholds
due to mass splittings at the high scale $M_{heavy}\neq M_{\rm GUT}$
arising from the infinite many 
massive string states \cite{lan93}.
In the following sections we shall discuss this question of string
unification in detail.

\section{Gauge coupling unification}

In heterotic superstring theories all couplings are related to 
the universal string coupling constant $g_{\rm string}$ at
the string scale 
$M_{\rm string}\sim 1/\sqrt{\al'}$, with $\al'$ being the 
inverse string tension. It is a free parameter which is fixed by the dilaton
vacuum expectation value $g_{\rm string}^{-2}=\fc{S+\ov S}{2}$.
In general this amounts to 
string unification, i.e. at the string scale \ms\ all gauge
and Yukawa couplings are proportional
to the string coupling and are therefore related to each other.
For the gauge couplings (denoted by $g_a$) we have \cite{gin87}:

\be\label{hyper}
g^{2}_ak_a=g_{\rm string}^2=\fc{\kappa^2}{2\al'}\ .
\ee
Here, $k_a$ is 
the Kac--Moody level of the group factor labeled by $a$.  
The string coupling $g_{\rm string}$ 
is related to the gravitational coupling constant
$\kappa^{2}$. In particular this means that string 
theory itself provides gauge coupling
and Yukawa coupling unification even in absence of 
a grand unified gauge group.

To make contact with the observable world
one constructs the field--theoretical low--energy limit of a 
string vacuum. This is achieved by integrating out
all the massive string modes corresponding to excited string states 
as well as states with momentum
or winding quantum numbers in the internal dimensions. 
The resulting theory then describes the
physics of the massless string excitations at low energies 
$\mu < M_{\rm string}$ in field--theoretical
terms.
If one wants to state anything about higher energy scales one has to
take into account \tc\ $\triangle_a(M_{\rm string})$ 
to the bare couplings $g_a(M_{\rm string})$
due to the infinite tower of massive string modes. They change
the relations \req{hyper} to:

\be
g_a^{-2}=k_ag_{\rm string}^{-2}+\fc{1}{16\pi^2}\triangle_a\ ,
\label{triangle}
\ee
The corrections in \req{triangle}
may spoil the string tree--level result \req{hyper} and 
split the one--loop gauge couplings at
$M_{\rm string}$.
This splitting could allow for an effective unification at a scale 
$M_{\rm GUT} <M_{\rm string}$ or destroy the unification.
 
The general expression of $\triangle_a$ for heterotic tachyon--free 
string vacua is given in \cite{vk}. Various contributions to $\triangle_a$ 
have been determined for several classes of models:
First in \cite{vk} for two $\Z_3$ orbifold models with a (2,2) 
world--sheet supersymmetry \cite{dhvw}. 
This has been extended to fermionic constructions
in \cite{fermionic}. Threshold corrections for (0,2) orbifold models 
with quantized Wilson lines \cite{hpn1} have been 
calculated in \cite{mns}.
Threshold corrections for the quintic threefold
and other Calabi--Yau manifolds \cite{chsw} with gauge group 
$E_6\times E_8$ can be found in \cite{ber1,kl2}.
In toroidal orbifold compactifications ~\cite{dhvw}
moduli dependent threshold corrections 
arise only from N=2 supersymmetric sectors. They have been
determined for some orbifold compactifications in 
\cite{DKL2}--\cite{cfilq} and for
more general orbifolds in \cite{ms1}. 
The full moduli dependence\footnote{A lowest expansion
result in the Wilson line modulus has been obtained in \cite{agnt2,clm2}.}  
of threshold corrections 
for (0,2) orbifold compactifications with continuous Wilson lines
has been first derived in \cite{ms4,ms5}.
These models contain continuous background gauge fields in addition
to the usual moduli fields \cite{hpn2}. In most of the cases
these models are (0,2) compactifications.
In all the above orbifold examples the threshold corrections $\triangle_a$ 
can be decomposed into three parts:

\be
\triangle_a=\tilde \triangle_a-b_a^{N=2}\triangle+k_a\ Y\ .
\label{form}
\ee
Here the gauge group dependent part is divided into two pieces:
The moduli independent part $\tilde \triangle_a$ containing
the contribution of the N=1 supersymmetric sectors as 
well as scheme dependent parts which are proportional to $b_a$.
This prefactor $b_a$ is related to the one--loop 
$\bet$--function: $\bet_a=b_ag_a^3/16\pi^2$.
Furthermore the moduli dependent part $b^{N=2}_a\triangle$ with
$b_a^{N=2}$ being related to the anomaly coefficient $b'_a$ by
$b_a^{N=2}=b_a'-k_a\de_{\rm GS}$.
The gauge group independent part $Y$ contains the gravitational 
back--reaction to the background gauge fields as well as other universal parts 
\cite{vk,dfkz,kl2,kk}. They are absorbed into the definition of 
$g_{\rm string}$: $g_{\rm string}^{-2}=\fc{S+\ov S}{2}+\fc{1}{16\pi^2}Y$. 
The scheme dependent parts are the 
IR--regulators for both field-- and string theory as well as
the UV--regulator for field theory. The latter is put into the definition of 
$M_{\rm string}$ in the $\ov{\rm DR}$ scheme \cite{vk}: 

\be
M_{\rm string}=2\fc{e^{(1-\ga_{\rm E})/2} 3^{-3/4}}{\sqrt{2\pi \al'}}= 
0.527\ g_{\rm string} \times 10^{18}\ {\rm GeV}\ .
\label{kaprel}
\ee
The constant of the string IR--regulator as well as 
the universal part due to gravity were
recently determined in \cite{kk}.

The identities \req{triangle} are the key to extract any string--implication 
for low--energy physics. They serve as boundary conditions for
our running field--theoretical couplings valid below \ms\ \cite{basicweinberg}.
Therefore they are the foundation of any discussion about both
low--energy predictions and gauge coupling unification.
The evolution equations\footnote{We neglect the N=1 part of 
$\tilde \triangle_a$ which is small compared to 
$b_a^{N=2}\triangle$ \cite{vk,fermionic,mns}.} valid below $M_{\rm string}$ 

\be
\fc{1}{g_a^2(\mu)}=\fc{k_a}{g_{\rm string}^2}+\fc{b_a}{16\pi^2} 
\ln \fc{M_{\rm string}^2}{\mu^2}-\fc{1}{16\pi^2}b^{N=2}_a\triangle\ ,
\label{running}
\ee
allow us to determine $\sin^2 \th_{\rm W}$ and
$\al_{\rm S}$ at $M_Z$. 
After eliminating $g_{\rm string}$ in the second and third equations 
one obtains

\bea\label{mz}
\sin^2\th_{\rm
W}(M_Z)&=&\ds{\fc{k_2}{k_1+k_2}-\fc{k_1}{k_1+k_2}\fc{\al_{em}(M_Z)}{4\pi}
\lf[\Ac\ln\lf(\fc{M_{\rm string}^2}{M_Z^2}\ri)-\Ac'\ \triangle\ri]\ ,}\nnn
\al_{S}^{-1}(M_Z)&=&\ds{\fc{k_3}{k_1+k_2}\lf[\al_{em}^{-1}(M_Z)-\fc{1}{4\pi}\Bc
\ln\lf(\fc{M_{\rm string}^2}{M_Z^2}\ri)+\fc{1}{4\pi}\Bc'\ \triangle\ri]\ ,}
\eea
with  $\Ac=\fc{k_2}{k_1}b_1-b_2, \Bc=b_1+b_2-\fc{k_1+k_2}{k_3}b_3$ and 
$\Ac',\Bc'$ are
obtained by exchanging $b_i\ra b_i'$. For the MSSM one has $\Ac=\fc{28}{5},
\Bc=20$.
However to arrive at the predictions of the MSSM \req{exp} 
one needs huge string 
threshold corrections $\triangle$ due to the large value of 
\ms\ \ $(3/5k_1=k_2=k_3=1)$: 

\be\label{tri}
\triangle=\fc{\Ac}{\Ac'}\lf[\ln\lf(\fc{M_{\rm string}^2}{M_{\rm
GUT}^2}\ri)+\fc{32\pi\de_{\sin^2\th_{\rm W}}}{5\Ac\al_{em}(M_Z)}\ri]\ .
\ee
At the same time, the N=2 spectrum of the underlying theory
encoded in $\Ac',\Bc'$ which enters the threshold corrections
has to fulfill the condition

\be\label{cond}
\fc{\Bc'}{\Ac'}=\fc{\Bc}{\Ac}\ \fc{\ln\lf(
\fc{M_{\rm string}^2}{M_{\rm GUT}^2}\ri)+\fc{32\pi}{3\Bc}
\de_{\al_{\rm S}^{-1}}}{\ln\lf(\fc{M_{\rm string}^2}{M_{\rm GUT}^2}\ri)+
\fc{32\pi}{5\Ac}\fc{\de_{\sin\th_{\rm W}^2}}{\al_{em}(M_Z)}}\ ,
\ee
where $\de$ represents the experimental uncertainties
appearing in \req{exp}. In addition
$\de$  may also contain SUSY thresholds.

For concreteness and as an illustration
let us take the $\Z_8$ orbifold example of \cite{IL} with
$\Ac'=-2,\Bc'=-6$ and $b_1'+b_2'=-10$.
It is one of the few orbifolds left over after 
imposing the conditions on target--space duality anomaly cancellation 
\cite{IL}.
To estimate the size of $\triangle$ one may take in eq. ~\req{kaprel}
$g_{\rm string} \sim 0.7$ corresponding to $\al_{\rm string}\sim\fc{1}{26}$,
i.e. $M_{\rm string}/M_{\rm GUT}\sim 20$. 
Of course this is a rough estimate since
 $M_{\rm string}$ is determined by the first eq. of \req{running} 
together with \req{kaprel}. Nevertheless, the qualitative picture does
not change.
Therefore to predict the correct low--energy parameter \req{mz}
eq. \req{tri} tells us that one needs threshold correction of considerable 
size:

\be
-17.1\leq\triangle\leq-16.3\ .
\label{size}
\ee

\section{String thresholds}

The  construction of a realistic unified string model boils down to the 
question of how to achieve thresholds of that size. To settle the question
we need explicit calculations within the given candidate string model.
There we can encounter various types of threshold effects. Some depend 
continuously, others discretely on the values of the moduli fields.
For historic reasons we also have to distinguish between thresholds
that do or do not depend on Wilson lines.
The reason is the fact that the calculations
in the latter models are considerably simpler and for some  time were
the only available results. They were then used to estimate the thresholds
in models with gauge group $SU(3)\times SU(2)\times U(1)$ and three
families, although as a string model no such orbifold can be constructed 
without Wilson lines.
Therefore, the really relevant thresholds are, of course, the ones found
in the (0,2) orbifold models with Wilson lines \cite{ms4}
which may both break the gauge group and reduce its rank.
We will discuss the various contributions within the
framework of our illustrative model. 
However the discussion can easily applied for all other orbifolds.
The threshold corrections depend on the $T$ and $U$ modulus
describing the size and shape of the internal torus lattice. In addition
they may depend on non--trivial gauge background fields encoded
in the Wilson line modulus $B$.

Moduli dependent threshold corrections $\triangle$ can be
of significant size for an appropriate 
choice  of the vevs of the background fields $T,U,B,\ldots$
which enter these functions. Of course in the decompactification 
limit $T\ra i \infty$ these corrections become always arbitrarily huge.
This is in contrast to fermionic string compactifications 
or N=1 sectors of heterotic superstring compactifications. 
There one can argue that {\em moduli--independent}
threshold corrections cannot become huge at all \cite{df}. This 
is in precise agreement with the results found earlier in 
\cite{vk,fermionic}.
In field theory threshold corrections can be estimated with the 
formula \cite{basicweinberg}

\be\label{field}
\triangle=\sum_{n,m,k}\ln\lf(\fc{M_{n,m,k}^2}{M_{\rm string}^2}\ri)\ ,
\ee
with $n,m$ being the winding and momentum, respectively and $k$
the gauge quantum number of all particles running in the loop.
The string mass in the $N=2$ sector of the $\Z_8$ model we consider
later with a non--trivial gauge background in the internal
directions is determined by \cite{ms5} :

\bea\label{mass}
\al'M_{n,m,k}^2&=&4|p_R|^2\nnn
p_R&=&\ds{\fc{1}{\sqrt{Y}}\lf[(\fc{T}{2\al'}U-B^2)n_2+\fc{T}{2\al'}n_1
-Um_1+m_2+Bk_2\ri]}\nnn
Y&=&\ds{-\fc{1}{2\al'}(T-\ov T)(U-\ov U)
+(B-\ov B)^2\ .}
\eea
In addition a physical state $|n,m,k,l\rng$ has to obey the modular 
invariance condition 
$m_1n_1+ m_2n_2+ k_1^2-k_1k_2+k_2^2-k_2k_3-k_2k_4+k_3^2+k_4^2
=1-N_L-\h l^2_{E_8'}$. Therefore the sum in \req{field}
should be restricted to these states. This also guarantees
its convergence after a proper regularization.
In \req{field} cancellations between the contributions of 
various string states may arise. E.g. at the critical point $T=i=U$ 
where all masses appear in integers of \ms\ such cancellations occur.
They are the reason for the smallness of the corrections 
calculated in \cite{vk,mns} and in all the fermionic models \cite{fermionic}.
Let us investigate this in more detail. 
The simplest case ($B=0$) for moduli dependent \tcgc\  was derived 
in \cite{DKL2} : 

\be
\triangle(T,U)=\ln\lf[\fc{-iT+i\ov T}{2\al'}\lf|\eta\lf(\fc{T}{2\al'}
\ri)\ri|^4\ri]+\ln\lf[(-iU+i\ov U)\lf|\eta(U)\ri|^4\ri]\ .
\label{22th}
\ee
Formula \req{22th} can be used for any toroidal orbifold compactifications,
where the two--dimensional subplane of the internal lattice
which is responsible for the N=2 structure factorize from the 
remaining part of the lattice. If the latter condition
does not hold, \req{22th} is generalized \cite{ms1}.

\bdm
\ba{|c|c|c|c|c|c|}\hline
&&&&&\\[-.25cm] 
\ &T/2\al'&U&M^2\al' &ln(M^2\al')&\Delta^{II}\\[-.25cm]
&&&&&\\ \hline &&&&&\\[-.25cm]
Ia&i&i     & 1  &0&-0.72\\[-.25cm]
&&&&&\\ \hline &&&&&\\[-.25cm]
Ib&1.25i&i&\fc{4}{5} &-0.22  &-0.76\\[-.25cm]
&&&&&\\ \hline &&&&&\\[-.25cm] 
Ic&4.5i&4.5i      &\fc{4}{81} &-3.01  &-5.03\\[-.25cm]
&&&&&\\ \hline &&&&&\\[-.25cm] 
Id&18.7i&i&\fc{10}{187} &-2.93&-16.3\\[-.25cm]
&&&&&\\ \hline
\ea
\edm
\begin{center}
{\em Table 1: Lowest mass $M^2$ of particles charged\\ under $G_A$
and threshold corrections $\triangle(T,U)$.}
\end{center}

In Table 1 we determine the mass of the lowest massive string state being 
charged under the considered unbroken gauge group $G_A$
and the threshold corrections $\triangle(T,U)$ for some values 
of $T$ and $U$.

The influence of moduli dependent \tc\ to low--energy physics [entailed
in eqs. \req{mz}]
has  until now only been discussed  for orbifold compactifications 
without Wilson lines by using \req{22th}.
In these cases the corrections only depend on the two moduli $T,U$.
However to obtain corrections of the size $\triangle\sim-16.3$ one would need
the vevs $\fc{T}{2\al'}=18.7, U=i$
which are  far away from the
self--dual points \cite{ILR,IL}. It remains an open question whether and 
how such big
vevs of $T$ can be obtained in a natural way in string theory.

A generalization of eq. \req{22th} appears when turning on non--vanishing
gauge background fields $B\neq0$.
According to \req{mass} the mass of the heavy string states now becomes
$B$--dependent and therefore also the \tc\ change.
This kind of corrections were recently determined in \cite{ms4}.
The general expression there is

\be\label{c12}
\triangle^{II}(T,U,B)=\fc{1}{12}\ln\lf[\fc{Y^{12}}{1728^4}
\lf|\Cc_{12}(\Om)\ri|^2\ri]\ ,
\ee
where $B$ is the Wilson line modulus, 
$\Om=\lf(\ba{cc} \fc{T}{2\al'}&B\\ -B&U\ea\ri)$ and $\Cc_{12}$ is a combination
of $g=2$ elliptic theta functions explained in detail in \cite{ms5}. 
It applies to gauge groups $G_A$
which are not affected by the Wilson line mechanism. The case where the gauge
group is broken by the Wilson line will be discussed later (those threshold
corrections will be singular in the limit of vanishing $B$).
Whereas the effect of quantized Wilson lines $B$ on \tc\ has already 
been discussed  in \cite{mns} the function 
$\Delta^{II}(T,U,B)$ now allows us to study the effect of a continuous variation
in $B$.

\vspace{1cm}
\epsfbox[-80 0 500 210]{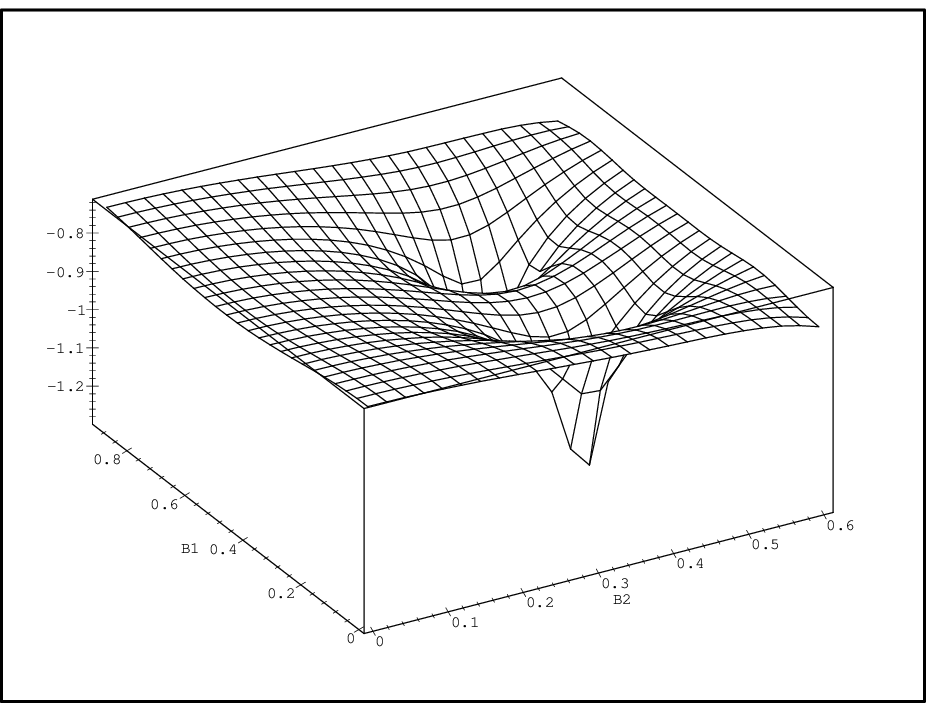}
\bc
{\small \em Fig.2 -- Dependence of the \tc\ $\triangle^{II}$\\
on the Wilson line modulus $B=B_1+iB_2$ for $\fc{T}{2\al'}=i=U$.}
\ec
We see in Fig.2  that the threshold corrections change very little
with the Wilson line modulus $B$. They are comparable
with $\triangle=-0.72$ corresponding to the case of $B=0$.
In this case eq. \req{c12} becomes eq. \req{22th} 
for $\fc{T}{2\al'}=i=U$.

So far all these calculations have been done within models 
where the considered gauge group $G_A$ is not broken by the Wilson line
and its matter representations are not projected out.
To arrive at SM like gauge groups with the matter content
of the MSSM one has to break the considered gauge group with 
a Wilson line.

From the phenomenological point of 
view \cite{wend}, the most promising 
class of string vacua is provided by
(0,2) compactifications equipped with a non--trivial gauge background in the 
internal space which breaks the $E_6$ gauge group
down to a SM--like gauge group \cite{w1,w2,hpn1,hpn2,stringgut}. 
Since the internal space is not simply connected, these gauge fields cannot be 
gauged away and may break the gauge group.
Some of the problems  present in (2,2) compactifications 
with $E_6$ as a  grand unified group like e.g. 
the doublet--triplet splitting problem, the fine--tuning problem and 
Yukawa coupling unification may be absent in (0,2) compactifications.
It is important that these properties can be studied in the full
string theory, not just in the field theoretic limit \cite{w1}.
The background gauge fields give rise to a new class of massless moduli
fields again denoted by $B$
which have quite different low--energy implications than the 
usual moduli arising from the geometry of the internal manifold itself.
In this framework the question of string unification can now be discussed
for realistic string models.
The threshold corrections for our illustrative model take the form \cite{ms4}

\be\label{chi10}
\triangle^I(T,U,B)=\fc{1}{10}\ln \lf[Y^{10}\lf|\fc{1}{128}
\prod_{k=1}^{10}\vartheta_k(\Om)\ri|^4\ri]\ ,
\ee
where $\vartheta_k$ are the ten even
$g=2$ theta--functions \cite{ms5}. Equipped with this result we can now
investigate the influence of the B--modulus on the thresholds and see how the
conclusions of ref. \cite{ILR,IL} might be modified. 
The results for a representative set of background vevs is displayed in
Fig.3.

\vspace{1cm}
\epsfbox[-80 0 500 210]{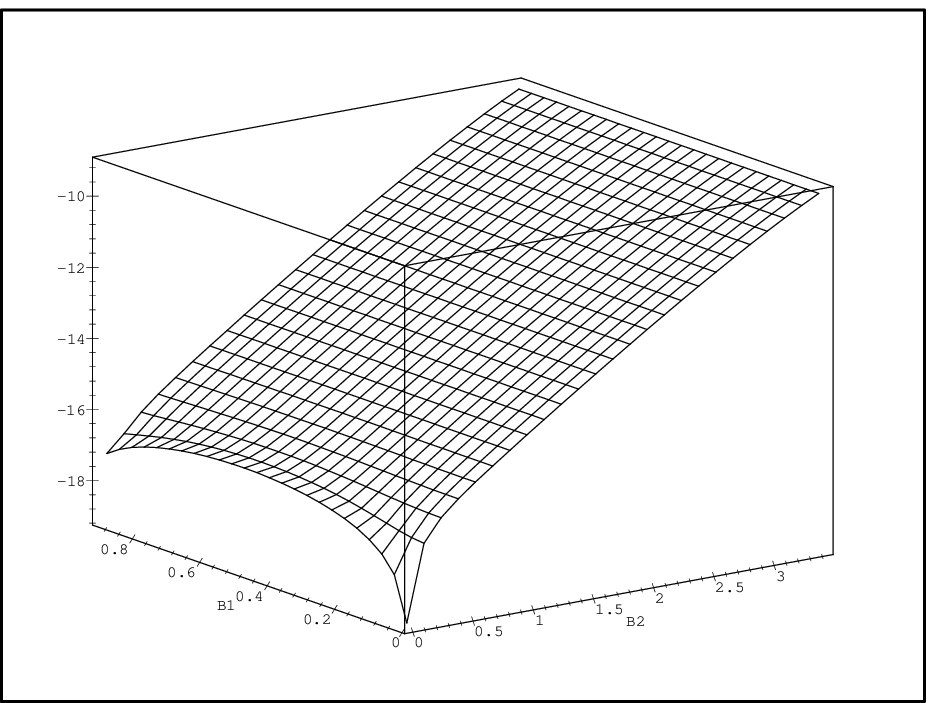}
\bc
{\small \em Fig.3 -- Dependence of the \tc\ $\triangle^I$ \\
on the Wilson line modulus $B=B_1+iB_2$ for $\fc{T}{2\al'}=4.5i=U$.}
\ec
From this picture we see that threshold corrections of
$\triangle\sim-16.3$ can be obtained for the choice of 
$\fc{T}{2\al'}\sim 4.5i\sim U$ and $B=\h$. This has to be compared to the model
in ref. \cite{IL} where such a value was achieved with $T=18.7i$ and $B=0$.
This turns out to be a general property of the models under
consideration. With more moduli, sizeable threshold effects are achieved
even with moderate values of the vevs of the background fields.

\section{String unification}

Equipped with these explicit calculations of string threshold
corrections we can now ask the question how string theory might lead
to the correct prediction of gauge coupling constants. We  also
hope to deduce information on the spectrum of theories that 
lead to successful gauge coupling unification. 

The modulus plays the r\^ole of an adjoint Higgs field which breaks 
e.g. the $G_A=E_6$ down to a SM like gauge group $G_a$. 
According to eq. \req{mass} the vev
of this field gives some particles masses between zero and \ms. 
This is known as the stringy Higgs effect. 
Such additional intermediate fields may be very important to generate 
high scale thresholds.
Sizeable \tc\ $\triangle$ can only appear if some particles have masses
 different from the string scale \ms\ and where cancellations
between different states as mentioned above do not take place. 
In particular some gauge bosons of $G_A$ become massive receiving the mass:

\be
\al'M_I^2=\fc{4}{Y}|B|^2\ .
\ee
As before let us investigate the masses of the lightest massive particles
charged under the gauge group $G_a$. For our concrete model we have 
$M_{\rm string}=3.6\cdot 10^{17}\gev$.

\bdm
\ba{|c|c|c|c|c|c|c|}\hline
&&&&&&\\[-.25cm] 
\ &T/2\al'&U&B&M_I\  [\gev] &ln(M_I^2\al') &\Delta^I\\[-.25cm]
&&&&&&\\ \hline &&&&&&\\[-.25cm]
IIa&i&i&\fc{1}{10^{5}}& 8.4\cdot 10^{12} &-23.0  &-10.03\\[-.25cm]
&&&&&&\\ \hline &&&&&&\\[-.25cm]
IIb&i&i&\h     &4.2\cdot 10^{17}     &-1.39           &-1.72\\[-.25cm]
&&&&&&\\ \hline &&&&&&\\[-.25cm] 
IIc&1.25i&i&\h&3.7\cdot 10^{17}  &-1.61 &-2.12  \\[-.25cm]
&&&&&&\\ \hline &&&&&&\\[-.25cm] 
IId&4i&i&\h     &2.1\cdot 10^{17} &-2.78    &-7.86\\[-.25cm]
&&&&&&\\ \hline &&&&&&\\[-.25cm] 
IIe&4.5i&4.5i&\h     &9.3\cdot 10^{16}&-4.39   &     -16.3\\[-.25cm]
&&&&&&\\ \hline &&&&&&\\[-.25cm]
IIf&18.7i&i&\h&1.1\cdot 10^{16}&-4.31            &-43.3\\[-.25cm]
&&&&&&\\ \hline
\ea
\edm
\begin{center}
\vspace{.3cm}
{\em Table 2: Lowest mass $M_I$ of particles charged\\ 
under $G_a$ and threshold corrections $\triangle^I$ for $B\neq0$.}
\end{center}
\ \\
Whereas $\Delta^{II}$ describes \tc\ w.r.t. to a gauge group 
which is not broken when turning on a vev of $B$, 
now the gauge group is broken for $B\neq 0$ and in particular
this means that the threshold $\triangle^I$ shows a logarithmic singularity
for $B\ra 0$ when the full gauge symmetry is restored. 
This behaviour is known from field theory and the effects of 
the heavy string states
can be decoupled from the former:
Then the part of $\triangle_a^I$ in \req{triangle} which is only due to
the massive particles becomes \cite{ms4,clm2}

\be
\fc{b_A-b_a}{16\pi^2}\ln\fc{M_{\rm string}^2}{|B|^2}
-\fc{b_A'}{16\pi^2}\ln\lf|\eta\lf(\fc{T}{2\al'}\ri)\eta(U)\ri|^4\ ,
\ee
where the first part accounts for the new particles appearing
at the intermediate scale of $M_I$
and the other part takes into account the contributions
of the heavy string states.
One of the questions of string unification concerns the size of this
intermediate scale $M_I$.
In a standard grand unified model one would be tempted to identify
$M_I$ with $M_{\rm GUT}$. While this would also be a possibility for
string unification, we have in string theory in addition the possibility
to consider $M_I>M_{\rm GUT}$. The question remains whether the 
thresholds in that case can be big enough, as we shall discuss in a moment.
Let us first discuss the general consequences of our
results for the idea of string unification without a grand unified
gauge group.
Due to the specific form of the threshold corrections
in eq. \req{running} unification always takes place if the condition 
$\Ac\Bc'=\Ac'\Bc$
is met within the errors arising from the uncertainties in \req{exp}. 
It guarantees that all three gauge couplings meet at
a single point $M_{\rm X}$ \cite{IL}:

\be
M_{\rm X}=M_{\rm string}\ e^{\h\fc{\Ac'}{\Ac}\triangle}\ .
\ee
For our concrete model this leads to $M_{\rm X}\sim2\cdot 10^{16}\gev$.
Given these results we can now study the relation between $M_I$ and
$M_{\rm X}$, which plays the r\^ole of the GUT--scale in string unified models.
As a concrete example, consider the model $IIe$ in Table 2. It leads
to an intermediate scale $M_I$ which is a factor 3.9 smaller than the
string scale, thus $\sim 10^{17}\gev$, although the apparent unification
scale is as low as $2\times 10^{16}\gev$. We thus have an explicit example
of a string model where all the non--MSSM particles are above
$9.3\cdot10^{16}\gev$, but still a correct prediction of the low energy 
parameters
emerges.
{\em Thus string unification can be achieved without the introduction of
a small intermediate scale.}

Of course, there are also other possibilities which lead to the 
correct low--energy predictions.
Instead of large threshold corrections one could consider
a non--standard hypercharge
normalization, i.e. a $k_1\neq 5/3$ \cite{ib}.
This would maintain gauge coupling unification at the string scale
{\em with} the correct values of $\sin^2\th_{\rm W}(M_Z)$ and 
$\al_{\rm S}(M_Z)$.
However, it is very hard to construct such models.
A further possibility  would be to give up the idea of
gauge coupling unification within the MSSM by introducing
extra massless particles such as $\bf{(3,2)}$ w.r.t.
$SU(3)\times SU(2)$ in addition to those of the SM \cite{inter,df}.
A careful choice of these matter fields may lead to sizable
additional intermediate threshold corrections in \req{mz} thus
allowing for the correct low--energy data \req{exp}.
Unfortunately the price for that is exactly an introduction of a new 
intermediate scale of $M_I\sim 10^{12-14}\gev$. It seems to be hard to
explain such a small scale naturally in the framework of string theory.
In some sense such a model can be compared to the model $IIa$ in table 2.
Other possible corrections to \req{mz} may arise
from an extended gauge structure between \mx\ and \ms.
However this might even enhance the disagreement
with the experiment \cite{df}. Finally a modification of \req{mz}
appears from the scheme conversion from the string-- or SUSY--based 
$\ov{DR}$ scheme
 to the $\ov{MS}$ scheme relevant for the low--energy physics data \req{exp}.
However these effects are shown to be small \cite{df}.

\section{Conclusions}

 We have seen that 
string unification is easily achieved with moduli
dependent threshold corrections within (0,2) superstring compactification.
The Wilson line dependence
of these functions is comparable to that on the $T$ and $U$ fields thus
offering the interesting possibility of large thresholds with
background configurations of moderate size. All non--MSSM like
 states can e.g. be
heavier than $1/4$ of the string scale, still leading to an apparent
unification scale of $M_{\rm X}=\fc{1}{20}M_{\rm string}$.
We do not need vevs of the moduli fields that are of the order 20 away
from the natural scale, neither do we need to introduce particles at
a new intermediate scale that is small compared to $M_{\rm string}$.
The situation could be even more improved with a higher number of
moduli fields entering the threshold corrections:
They may come from other orbifold planes giving rise to N=2 sectors or 
from additional Wilson lines.
We think that
the actual moderate vevs of the underlying moduli fields can be fixed 
by non--perturbative effects as e.g. gaugino condensation. 

In this mechanism of supersymmetry breakdown we have seen the central
role played by the dilaton. Fixing the vev of the dilaton to a
satisfactory value is possible in the presence a nonperturbative modification
of the gauge kinetic function that seems to arise in
a broad class of string theories. It also leads to a solution of a
longstanding problem concerning the overall size of the gauge coupling
constant, since it allows weak coupling even for $<S>=1$.

\section{Acknowledgements}

We would like to thank Z. Lalak, P. Mayr, A. Niemeyer and
S. Stieberger  for helpful discussions
and comments. This work was partially supported
by Deutsche Forschungsgemeinschaft SFB-375-95 (research in astroparticle
physics) as well as EU grants SC1-CT91-0729 and SC1-CT92-0789.

\end{document}